\documentclass[preprint,showpacs,preprintnumbers,amsmath,amssymb]{revtex4}
\def\beq{\begin{equation}}
\def\eeq{\end{equation}}
\def\beqa{\begin{eqnarray}}
\def\eeqa{\end{eqnarray}}
\def\l{\left}
\def\r{\right}
\def\bdi{\begin{displaymath}}
\def\edi{\end{displaymath}}

\usepackage{graphicx}
\usepackage{dcolumn}
\usepackage{bm}
\begin{document}

\preprint{APS/123-QED}
 
\title{Mechanical unfolding of directed polymers in a poor solvent: novel critical exponents}
\author{A. Rosa$^1$, D. Marenduzzo$^{2}$, A. Maritan$^{1,3}$, F. Seno$^{4}$} 
\affiliation{
$^1$ International School for Advanced Studies (SISSA) and INFM, Via Beirut 2-4, 34014 Trieste, Italy\\ 
$^2$ Department of Physics, Theoretical Physics, University of Oxford, 1 Keble Road, Oxford OX1 3NP, England\\ 
$^3$ The Abdus Salam International Center for Theoretical Physics (ICTP), Strada Costiera 11, 34100 Trieste, Italy\\ 
$^4$ INFM and Dipartimento di Fisica - Universit\`a di Padova, Via Marzolo 8, 35131 Padova, Italy\\
}

\begin{abstract}
We study the thermodynamics of an exactly solvable model of a self-interacting partially directed self-avoiding walk (DSAW) in two dimensions, when a force is applied on one end of the chain. The critical force for the unfolding is determined exactly, as a function of the temperature, below the $\Theta$-transition. The transition is second order and characterized by new critical exponents which are determined by a careful numerical analysis. The usual polymer critical index $\nu$ on the critical line, and another one, which we call $\zeta$, take a non-trivial value which is numerically close to $2/3$.
\end{abstract}


\maketitle

\begin{section}{Introduction}\label{intro}

The nature of the collapsed phase that a polymer attains in poor solvent conditions is still under debate (see e.g. \cite{domb,owczarek2} and references therein). Recent experiments on pulling of polymers and biopolymers (see e.g. \cite{exp1,exp2,exp3}) have enhanced the theoretical interest on the unfolding transition a collapsed polymers undergoes when subjected to an external force, $f$, applied at its extrema. Until very recently, most of the existing studies on this subject dealt with refined version of the mean field studies originally proposed in Ref. \cite{halperin}. A common characteristic of such studies is that, for a self-attracting polymer, they predict a first order phase transition in any dimension at a critical force $f_c(T)$. At temperatures below the $\Theta$-transition, where the self-attraction prevails, and for an applied force less than $f_c$ the polymer is in a compact phase. For forces greater than $f_c$ the self-attraction is unable to maintain the polymer in its compact conformation and the polymer chain is stretched along the force direction. However in $d=2$ extensive Monte-Carlo simulations \cite{grass}, performed on a self-avoiding walk (SAW) model, suggested that the transition is second order. An exactly solvable model, on a lattice of fractal dimension $2$, has been analyzed in Ref. \cite{jpa} and a second order transition was found at a critical force $f_c(T)$. In Ref. \cite{condmatstretch} a rationale was given for the change in order of the transition as the spatial dimension, $d$, goes past $2$, by means of a renormalization group based argument. Within that framework, it was found that, near criticality, the projection of the end-to-end distance along the force direction per monomer goes like $f-f_c$, near the phase transition, where $f$ is the force and $f_c$ is the critical force. Numerical uncertainties are too big to critically test this prediction in the self-avoiding walk (SAW) model of Refs. \cite{grass,condmatstretch}. Furthermore it was predicted \cite{jpa,condmatstretch} that above two dimension the transition line $f_c(T)$ shows a re-entrance at low temperature, i.e. $f_c(T)$ increases at low $T$ and after reaching a maximum it decreases becoming zero at $T_{\theta}$, the $\Theta$-transition. This behaviour is similar to the one found in theoretical models of pulling of double strand DNA where, however, the re-entrance was found at all $d$ \cite{marenduzzo1,marenduzzo2,marenduzzo3}. Here we will consider a simplified polymer model where the chains are represented by partially directed walks, i.e. steps with negative projection along the $x-$axis, $(1,0)$, are forbidden. This model proved to be helpful  in the past in order to find the phase diagram in the (temperature, fugacity) plane for a simplified $\theta$ transition \cite{foster,yeomans,brak,prellberg,owczarek1}. We take advantage of previous contributions and generalize the model to the presence of a pulling force along the direction $(1,0)$ (see also Ref. \cite{zhou}). Surprisingly in this version the critical force as a function of the temperature $T$ can be found analytically. With transfer matrix techniques we find that the end-to-end distance per monomer goes like $(f-f_c)^{1/\zeta-1}$, with $\zeta < 1$. With a sophisticated enumeration technique \cite{prellberg} we show the correlation critical exponent, $\nu$, takes on a non trivial value on the critical line, numerically very close to $\nu_{\theta} = 2/3$, the exponent at the $\Theta$ transition. It is not clear whether this is an accidental degeneracy or if it can apply also in the undirected case, too. For example in the $3d$ Sierpinski gasket an exact renormalization leads to a non trivial $f$-dependence of $\nu$ \cite{jpa}).

Our work is structured as follows. In Section \ref{model}, we introduce the model and the basic quantities of interest. In Section \ref{tm}, we outline how the transfer matrix can be applied to our model, find explicitly the phase diagram (critical line) and give a rough estimate of the exponent $\zeta$. A scaling argument is proposed to suggest that at criticality $\nu=\zeta$. In Section \ref{methods}, we review the enumeration technique proposed in Ref. \cite{prellberg}, which we use in Section \ref{results} in order to estimate the value of $\nu$ on the critical line. In Section \ref{scaling}, we critically analyze our scaling ansatz and the hypothesis that $\nu=\zeta$. Finally, in Section \ref{conclusions} we draw our conclusions.
\end{section}

\begin{section}{The model}\label{model}

The model is a DSAW on a two-dimensional square lattice (see Fig. \ref{configurations}), with (non-consecutive) nearest-neighbor interactions. A force ${\mathbf f}$, directed along the same axis of the walk,  is pulling on one end of the DSAW, the other one being fixed. Given a particular configuration $\mathcal{C}$ the energy is
\beq \label{energy}
E_{\mathcal{C}} = - \epsilon m - f R_x,
\eeq
where $m$ is the number of interacting pairs and $\epsilon$ the energy per pairs, $f$ the modulus of the applied force and $R_x$ the longitudinal extension of the walk. Then, the canonical partition function can be written
\beq \label{canpartfunc}
{\mathcal Q}_L = {\mathcal Q}_L(\beta\epsilon, \beta f) =
\sum_{\mathcal{C}} e^{-{\beta} E_{\mathcal{C}}},
\eeq
where $L$ is the number of the steps of the walk and $\beta^{-1}=T$, is the temperature in units of the Boltzmann constant. From now on will set $\epsilon = 1$ without loss of generality.
\begin{figure}
\includegraphics[width=2.5in]{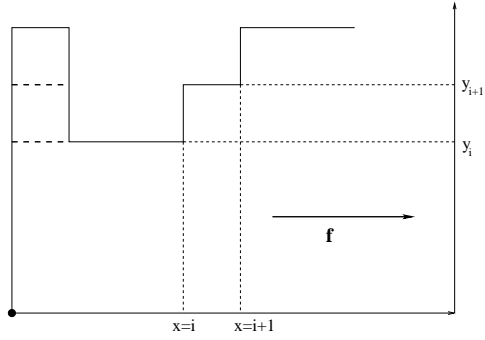}
\caption{An example of DSAW configuration. The quantities $y_i$ and the force direction are also displayed. Thick dashed lines indicate contacts.} \label{configurations}
\end{figure}
From the canonical partition function we construct the grancanonical partition function (generating function),
\beq\label{genfunc}
{\mathcal G}(T, f, z) = \sum_{L = 1}^{\infty}
{\mathcal Q}_L z^L,
\eeq
$z$ being the step fugacity. The (real) singularity closest to the origin, $z_c(T, f)$, of the generating function Eq. (\ref{genfunc}) is related to the free energy per monomer as follows
\beq \label{zcrit}
z_c(T, f) = \lim_{L\to\infty}\frac{\ln{{\mathcal Q}_L}}{L}
\eeq
From the singularities of the generating function when $f = 0$, a complete phase diagram can be extracted (see the next Section and also Refs. \cite{yeomans,brak,owczarek1}). In particular, a discontinuity is found in the free energy at a value of $T=T_{\theta}$, called the $\Theta$ temperature.

\end{section}

\begin{section}{Transfer matrix calculations and phase diagram}\label{tm}

Starting from the definition of the generating function, Eq. (\ref{genfunc}), when $f=0$, we observe that it can be conveniently rewritten \cite{foster,yeomans} as
\beq \label{partitions}
{\mathcal G}(T,f,z)=\sum_{L_x}
{\mathcal G}_{L_x}(T, z) \exp{\l(\beta fL_x\r)},
\eeq
where ${\mathcal G}_{L_x} = \sum_L {\mathcal Q}_{L,L_x}z^L$, ${\mathcal Q}_{L,L_x}$ being the partition function restricted to walks of total length $L$ of which  $L_x$ steps are along the $x$ direction.

This is useful because now ${\mathcal G}_{L_x}$ can be written in terms of a transfer matrix, $T$, of dimensionality $L_y^2$, where $L_y$ is the size of our system along the $y$-direction\cite{foster,yeomans}. Such a transfer matrix $T$ is defined via its actions on the vectors $\l\{v_i\r\}_{i=1,\ldots,L_y^2}$, with $v_i=(y_i,y_{i+1})$, $y_i$ being the height of the site in the $i-$th row which precedes the right-bound horizontal link in that column (see Fig. \ref{configurations}), as follows:
\beqa \label{tm_definition}
T(v_i,v_{i+1})& =
&\exp[{\beta\epsilon\l(\min\{|r_i|,|r_{i+1}|\}\r)}\theta(-r_i \cdot
r_{i+1})]\nonumber\\ & & \times\exp[(|r_i|+1)\ln{z}].\nonumber\\
\eeqa
where $r_i = y_{i+1}-y_i$ and $\theta(x)$ is the Heaviside step function. It can be shown that ${\mathcal G}(T, f=0, z)$ develops a singularity when $\lambda$, the largest eigenvalue of $T$, goes through $1$ \cite{derrida}. This means that for large $L_x$
\beq \label{transfer2}
{\mathcal G}_{L_x}\propto
(\lambda(T,z))^{L_x}.
\eeq
Consequently the force-dependent singularity, $z_c(T,f)$ occurs when
\beq \label{forcesingularity}
\lambda(T,z)\exp{\l(\beta f\r)}=1.
\eeq
Eq. (\ref{forcesingularity}) has a rather deep consequence. In order for the critical fugacity and hence the free energy to display a singularity at a non-zero value of the force, i.e. in order for the force induced unfolding transition to exist as a thermodynamic transition and not only as a crossover, it is necessary that $\lambda(T,z)$, the largest eigenvalue of the transfer matrix when there is no force, has itself a discontinuity as $z$ approaches $z_c(T,f=0)\equiv z_0$. Otherwise, from Eq. (\ref{forcesingularity}) it is clear that there can be no such singularity. If there is a transition, then we get the following equation for the critical force:
\beq \label{critforce}
f_c(T)=-T \lim_{z\to z_0^{-}} \ln{\lambda(T, z)}.
\eeq
In Eq. (\ref{critforce}) the value of $\lambda$ to be put in the right hand side of the equation is the one pertaining to the infinite system. $\lambda(T,z)$ for $z$ slightly less than $z_0$ is plotted in Fig. \ref{largesteigenvalue} with a lateral size $L_y$ up to $40$. It is rather clear that a singularity has to be expected at $z=z_0$ in the infinite size limit. It was indeed shown \cite{yeomans,brak} that for $T<T_{\theta}$ ($T_{\theta}=0.8205\ldots$ in this model) there is a singularity of the grand partition function for $z=z_0=\exp(-\beta)$ and for this value of the fugacity the biggest eigenvalue is strictly smaller than $1$, being:
\beq \label{eigenvalue}
\lambda(\beta,z=z_0=\exp(-\beta))\equiv\lambda(\beta) = \frac{z_0\l(1+\sqrt{z_0}\r)}{1-\sqrt{z_0}}.
\eeq
The $\Theta$ transition temperature is obtained when $\lambda(\beta,z=z_0=\exp(-\beta)) = 1$. Consequently, the critical line, $f_c(T)$, is obtained by putting $\lambda(\beta)=\exp{\l(-\beta f_c(T)\r)}$, i.e.:
\beq \label{critical_force}
f_c(T)=T\ln{\l[\frac{1-\exp{\l(-\beta/2\r)}} {\exp{\l(-\beta
\r)}\l(1+\exp{\l(-\beta/2\r)}\r)}\r]},
\eeq 
and is plotted in Fig. \ref{phase_diagram}, where also the results obtained with the transfer matrix with system size up to $L_y=40$ are displayed.
\begin{figure}
\includegraphics[width=2.5in,angle=270]{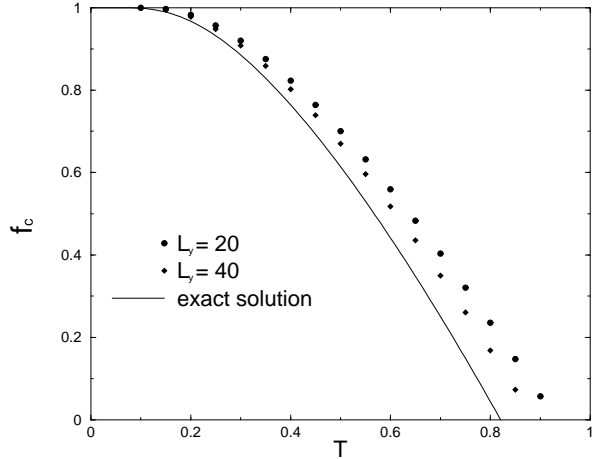}
\caption{Exact critical line as in Eq. (\ref{critical_force}) together
with points corresponding to estimates with the transfer matrix
calculation, with strip size $L_y=20,40$.} \label{phase_diagram}
\end{figure}
In view of Eqs. (\ref{forcesingularity}) and (\ref{critforce}), we can define a new critical exponent $\zeta$ which characterizes the directed self-avoiding walk. From Eq. (\ref{critforce}), if the largest eigenvalue approaches its limit value according to the law:
\beq \label{deltadef}
\lambda(z_0^{-})-\lambda(z)\sim(z_0-z)^{\zeta}
\eeq
then one straightforwardly obtains (via Eqs. (\ref{forcesingularity}) and (\ref{critforce}))
\beq\label{zeta_def}
\lim_{L\to\infty}\frac{\langle{R_x(L)}\rangle}{L}
\sim(f-f_c(T))^{1/\zeta- 1}
\eeq
where $\langle R_x(L)\rangle$ is the average projection of the end-to-end distance of the DSAW along the axis $(1,0)$. From Fig. \ref{largesteigenvalue} we estimated  $1/2<\zeta<1$, with $\zeta\simeq 0.7$ though a precise determination is difficult. If $\zeta<1$ in particular the transition is second order. It is widely accepted that for $d>2$ the transition is first order and so $\zeta=1$. In Ref. \cite{condmatstretch} a renormalization group based argument in $d=2$ on the other hand gave for the (undirected) SAW $\zeta=1/2$. This argument would apply also to the present case.
\begin{figure}
\includegraphics[width=2.5in,angle=270]{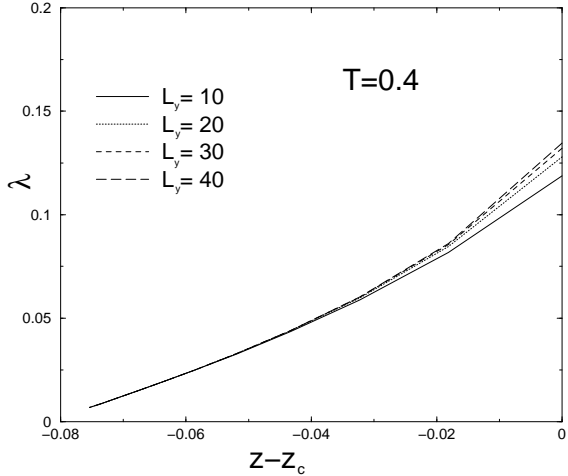}
\caption{Plot of the largest eigenvalue ($T=0.4<T_{\theta}$) versus $z$. It is apparent
that the largest eigenvalue approach a limit value as $z$
approaches $z_0 = \exp(-\beta)$ ($<1$) from below. Thus a transition
exists in the thermodynamic sense.} \label{largesteigenvalue}
\end{figure}
Given that the transition is second order in our model, it is also sensible to look for the value of the critical exponent $\nu$ (defined as $R_g\sim L^{\nu}$, for large number of steps $L$, where $R_g$ is the gyration radius of the $L-$site polymer. In Section \ref{scaling}, using a scaling argument, we shall demonstrate that $\nu = \zeta$.

In the next Section, we shall study the complete canonical partition function, Eq. (\ref{canpartfunc}), using a powerful method of exact enumeration introduced in \cite{prellberg}, that allows us to reach large values of $L$.

\end{section}

\begin{section}{The method of enumeration} \label{methods}
As already said, the configurations of the model are directed walks on a two-dimensional square lattice with nearest-neighbor interactions. For convenience, we demand that these walks end with a horizontal segment. Since the walks are directed in the $x$-direction we can describe these configurations through the distance $r_i$ between two horizontal steps, measured in the positive $y$ direction. Thus, we associate to each configuration an $N$-tuple $(r_1, r_2, ..., r_N)$ corresponding to a configuration of total length $L = \sum_{i=1}^N |r_i| + N$.

The energy due to the nearest-neighbor interactions for each of these configurations is (see Eq. (\ref{tm_definition}))
\beq \label{nninter}
U(r_1, r_2, ..., r_N) =
-\sum_{i=1}^{N-1}\min(|r_i|, |r_{i+1}|)\theta(-r_i \cdot r_{i+1}).
\eeq

In the following, we assign weights $x$ for steps in the horizontal direction and $y$ for steps in the vertical direction. Then, the canonical partition function is
\beqa \label{canpartfunc1}
{\mathcal Q}_L(x, y, \omega) & = & \sum_{N=1}^{L} (xe^{\beta f})^N
\nonumber\\ & \times &
\sum_{|r_1|+|r_2|+...+|r_N|=L-N}y^{L-N}\omega^{U(r_1, r_2, ..., r_N)},
\nonumber\\
\eeqa
where $\omega = \exp(\beta)$.

Now, it is convenient to consider the partition functions ${\mathcal Z}_L^{(r)} = {\mathcal Z}_L^{(r)}(x, y, \omega)$ for walks of total length $L+1$ which start with a vertical segment of height $r$. Then, we have 
\beq\label{canpartfunc2} 
{\mathcal Q}_{L+1}(x, y, \omega) = \sum_{r=-L}^{L} {\mathcal Z}_L^{(r)}, 
\eeq
(note that ${\mathcal Z}_L^{(0)} = x{\mathcal Q}_L(x, y, \omega)$) which satisfies the following recursion relation
\beq \label{recrelat}
{\mathcal Z}_L^{(r)} = xy^{|r|}\l\{ \delta_{|r|, L} + e^{\beta f}\sum_{s=-L+|r|+1}^{L-|r|-1} \omega^{U(r, s)} {\mathcal Z}_{L-|r|-1}^{(s)}\r\},
\eeq
obtained concatenating these walks. In the Eq. (\ref{recrelat}), $r = -L, ..., L$, with $L = 0, 1, 2, ...$. Using the symmetry ${\mathcal Z}_L^{(r)} = {\mathcal Z}_L^{(-r)}$, the Eq. (\ref{recrelat}) can be written only for non negative $r$ as
\beqa \label{recrelat1}
{\mathcal Z}_L^{(r)}
& = & xy^{r}\l\{ \delta_{r, L} + e^{\beta f}\sum_{s=0}^{L-r-1} {\mathcal Z}_{L-r-1}^{(s)} + \r. \nonumber\\
& & \l. + e^{\beta f}\sum_{s=1}^{L-r-1} \omega^{\min(r, s)} {\mathcal Z}_{L-r-1}^{(s)}\r\}.
\eeqa

Setting $x=y=1$ in (\ref{recrelat1}), we obtain the iteration scheme and the free energy ${\mathcal F}_L(\omega) = -\frac{1}{\beta L}\ln{\mathcal Z}_L^{(0)}$. The average longitudinal length of the walk $\langle R_x(L) \rangle$ is simply
\beq \label{xlength}
\langle R_x(L) \rangle = \frac{\partial}{\partial (\beta f)}\ln{\mathcal Z}_L^{(0)}.
\eeq

Then, we shall proceed as follows:
\begin{enumerate} 
\item we calculate the free energy using the iteration scheme proposed in Eq. (\ref{recrelat1}); 
\item using the Eq. (\ref{xlength}), we determine how the quantity $\langle R_x(L) \rangle/L$ varies against the applied force $f$. 
\end{enumerate}
\end{section}

\begin{section}{Results}\label{results}

The plot of $\langle R_x(L) \rangle/L$ vs. $f$ for various values of $L$ is represented on Fig. \ref{rvsf} at $T = 0.4$ which is below the $\Theta$ transition occurring at $T_{\theta} \simeq 0.8205...$. From Eq. (\ref{critical_force}) we have $f_c (T=0.4) \simeq 0.764...$

\begin{figure}
\includegraphics[width=2.5in, angle=270]{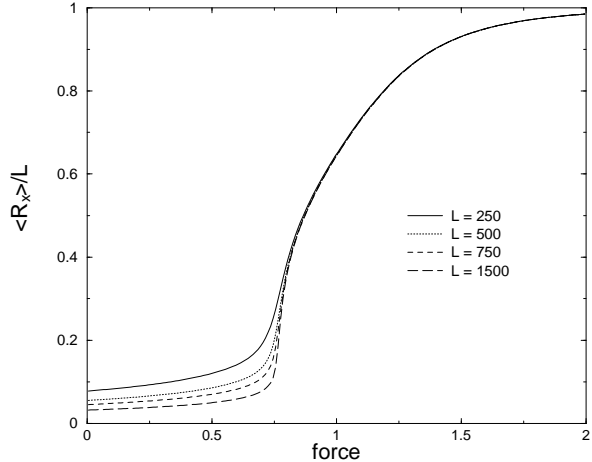}
\caption{Plot of $\langle R_x(L) \rangle/L$ vs. $f$ for various
length, $L$, of the walk.} 
\label{rvsf}
\end{figure}

From a careful examination of Fig. \ref{rvsf}, we deduce that the quantity $\langle R_x(L) \rangle/L$ decreases as $L^{\nu-1}$, where the {\it critical exponent} $\nu$ might depend on the temperature $T$. In particular the data are consistent with $\nu<1$ if $f\leq f_c(T)$ and $\nu=1$ if $f>f_c(T)$. In order to find more precise values for the critical exponent, we shall proceed along the same lines of Ref. \cite{prellberg}.

An estimation of the critical exponent through the use of the Pad{\'e} approximants \cite{guttmann} is given in Table \ref{pade}. 
\begin{table} 
\begin{tabular}{|*{3}{c|}}
\hline
{$f$} & {$\nu$} & {$L$}\\ 
\hline
\hline
{} & {} & {}\\
$< f_c$ & $ 0.501(7) $ & $ \le 1900 $\\
{} & {} & {}\\
$= f_c$ & $ 0.68(8) $ & $ \le 1900 $\\
{} & {} & {}\\
$> f_c$ & $ 1.00000 $ & $ \le 1300 $\\
{} & {} & {}\\
\hline
\end{tabular}
\caption{\label{pade}Estimates for the critical exponents from a Pad{\'e} approximants analysis. Note that in the $f>f_c$ case the error is completely negligible.}
\end{table}

Our estimate of the critical exponent, at the critical force, is close to $\frac{2}{3}$ , the $\nu$ value at the $\Theta$ point at $f=0$ \cite{brak,owczarek1}.

To get a deeper insight, let us define an $L-$dependent critical exponent $\nu(L)$ through the formula
\beq\label{critexp}
\nu(L) = \frac{\ln \langle R_x(L)(L+1) \rangle
 -\ln \langle R_x(L)(L)\rangle}{\ln(L+1)-\ln L}
\eeq
Plotting $\nu(L)$ versus an estimated correction-to-scaling term a
careful extrapolation to $L\rightarrow \infty$ can be performed,
determining the critical exponent $\nu$ for all the values of the
force. Let us consider three different regimes:

\begin{enumerate}

\item $f<f_c$.

As an example let us consider $f=0.4$. We have found that successive estimates for the exponent $\nu-1$ with increasing $L$ follow a straight line when plotted against a correction-to-scaling term of $1/L^{0.5}$ (see also the case of Ref. \cite{prellberg} at $f=0$). The plot is shown in Fig. \ref{nuvsL_down}. 
\begin{figure}
\includegraphics[width=2.5in, angle=270]{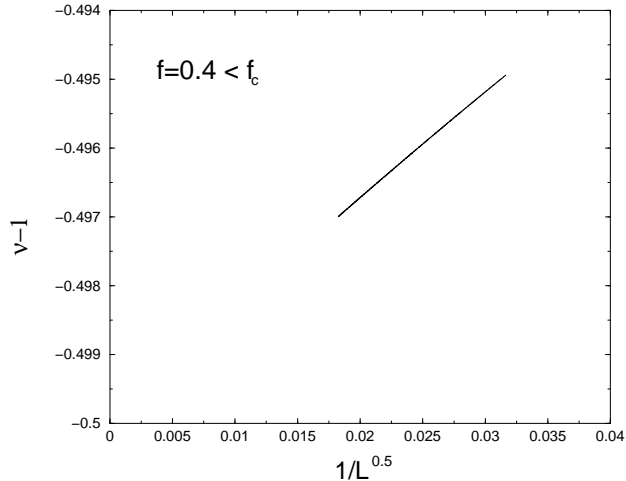}
\caption{Plot of $\nu-1$ vs. $1/L^{0.5}$, for $f<f_c$ and
$L$ up to 3000.} 
\label{nuvsL_down} 
\end{figure} 
The extrapolated value for $L \rightarrow \infty$ gives $\nu-1 \simeq -0.4998$, then $\nu \simeq \frac{1}{2}$, the exponent typical of a compact phase.

\item $f>f_c$.

As before, we have plotted the exponent $\nu-1$ against a well-defined correction-to-scaling term. Now, this term is order of $1/L$. Fig. \ref{nuvsL_up} shows the $f=1.0$ case as typical example.
\begin{figure}
\includegraphics[width=2.5in, angle=270]{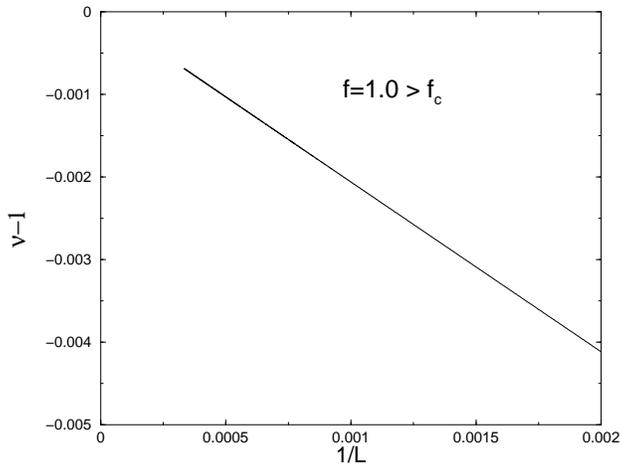}
\caption{Plot of $\nu-1$ vs. $1/L$, for $f>f_c$ and $L$ up
to 3000.} 
\label{nuvsL_up}
\end{figure}
Now, the extrapolated value gives $\nu-1 \simeq 3.0\times 10^{-6}$, then $\nu = 1$ within the numerical precision.

\item $f=f_c$.

Now, the correction-to-scaling term is of $1/L^{0.28}$ (see Fig. \ref{nuvsL_fcrit}) and $\nu-1 \simeq -0.3336$, which implies $\nu \simeq \frac{2}{3}$.
\begin{figure}
\includegraphics[width=2.5in, angle=270]{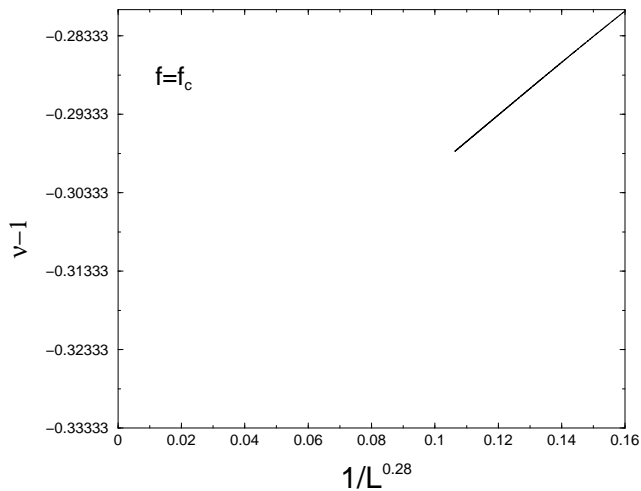}
\caption{Plot of $\nu-1$ vs. $1/L^{0.28}$,
for $f=f_c$ and $L$ up to 3000.}
\label{nuvsL_fcrit}
\end{figure}

\end{enumerate}

Thus we find that the value of $\nu$ for $f=f_c(T)$ is equal (in the limit of numerical precision) to the value $\frac{2}{3}$, which corresponds to that for $f=0$ at $T=T_{\theta}$ \cite{brak,owczarek1}. This is a non trivial result. In particular, we have to expect that along all the critical line $f=f_c(T)$, $\nu$ takes the value $\frac{2}{3}$. Moreover, as pointed out in \cite{owczarek1}, the correction-to-scaling term when $f=0$ and $T=T_{\theta}$ is of $1/L^{1/3}$. In our case, we found that this correction increases to $1/L^{0.28}$, for $T=0.4$. Within the numerical errors, this implies that the correction to scaling scaling term depends on force as well as on temperature and that the size of our system has to increase in order to find the right critical exponent.

Finally, in Table \ref{correctscal} we have summarized the above results.
\begin{table}
\begin{tabular}{|*{3}{c|}}
\hline
{$f$} & {$\nu$}\\
\hline
\hline
{} & {}\\
$< f_c$ & $1/2$ \\
{} & {}\\
$= f_c$ & $2/3$ \\
{} & {}\\
$> f_c$ & 1 \\
{} & {}\\
\hline
\end{tabular}
\caption{\label{correctscal}Estimates for the critical exponents from an extrapolation to
$L \rightarrow \infty$, obtained plotting the $L$-dependent critical
exponent $\nu(L)$, Eq. (\ref{critexp}), versus an estimated
correction-to-scaling term.}
\end{table}

In the next Section, we shall introduce a scaling theory that rationalizes for what we have found on the critical behaviour of the average horizontal end-to-end displacement $\langle R_x(L) \rangle$.
\end{section}

\begin{section}{Scaling theory}\label{scaling}
Our previous results suggest the following scaling ansatz (see also Ref. \cite{grass}):
\beq \label{scalfunc1}
\langle R_x(L) \rangle = L^{\nu}\Phi(\Delta f L^{\psi})
\eeq
where $\Delta f \equiv f-f_{c}(T)$. The scaling function $\Phi(x)$ must have the following behaviour:
\beq\label{phi}
\Phi(x) \sim \l\{
\begin{array}{lcc}
x^{(1-\nu)/\psi} & \mbox{if} & x \rightarrow +\infty \\
{} & {} & {}\\
\Phi_0 & \mbox{if} & x \rightarrow 0 \\
{} & {} & {}\\
|x|^{(1/2-\nu)/\psi} & \mbox{if} & x \rightarrow -\infty
\end{array}
\r.,
\eeq
with $\Phi_0$ a {\it not}-zero constant value. Then, the quantity $\langle R_x(L) \rangle$ obeys to the equations:
\beq\label{scalfunc2}
\langle R_x(L) \rangle \sim \l\{
\begin{array}{lccc}
L {\Delta f}^{(1-\nu)/\psi} & \mbox{if} & \Delta f > 0, & \frac{1-\nu}{\psi} > 0\\
{} & {} & {} & {} \\
L^{\nu} & \mbox{if} & \Delta f = 0 & {} \\
{} & {} & {} & {} \\
L^{1/2} |\Delta f|^{(1/2-\nu)/\psi} & \mbox{if} & \Delta f < 0, & \frac{1/2-\nu}{\psi} < 0
\end{array}
\r.,
\eeq
in agreement with the results found in the previous Section.

Now, let us observe that the free energy contribution to the singular part is
\beqa\label{singpart}
\Delta F & = & \langle R_x(L) \rangle f - \langle R_x(L) \rangle f_c = \langle R_x(L) \rangle \Delta f \nonumber\\
& = & \Delta f L^{\nu}\Phi(\Delta f L^{\psi})
\eeqa
where we have used Eq. (\ref{scalfunc1}), $f$ is the applied force and $\tilde{\Phi}(x) = x \Phi (x)$. Since $\Delta F$ is a contribution to the {\it total} free energy ({\it not} a free energy density), we expect it depends only on the ``dimensionless'' combination of the scaling fields $\Delta f$ and $L$ with the appropriate exponents. This implies that $\nu = \psi$. 

Comparing Eqs. (\ref{zeta_def}) and (\ref{scalfunc2}), we deduce that $\zeta = \nu = 2/3$. Then, Eq. (\ref{scalfunc1}) becomes $\langle R_x(L) \rangle = L^{2/3}\Phi(\Delta f L^{2/3})$.

To test this prediction, we have plotted in Fig. \ref{find_fcrit} the function $L^{-2/3}\langle R_x(L) \rangle$ versus $f-f_c$, where $f_c$ is again determined from the exact formula, Eq. (\ref{critical_force}). It is evident that, apart from obvious finite size scaling corrections, our ansatz is justified.
\begin{figure}
\includegraphics[width=2.5in, angle=270]{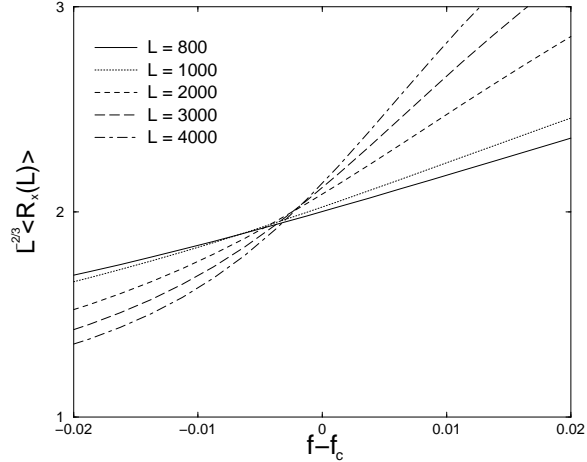}
\caption{Plot of $\langle R_x(L)\rangle$ scaled by $L^{2/3}$ versus $f-f_c$. Finite size scaling corrections to the critical force are evident.}
\label{find_fcrit}
\end{figure}
Then, we have derived the scaling function $\Phi(x)$. The final result is shown in Fig. \ref{scal_fig}.  
\begin{figure}
\includegraphics[width=2.5in, angle=270]{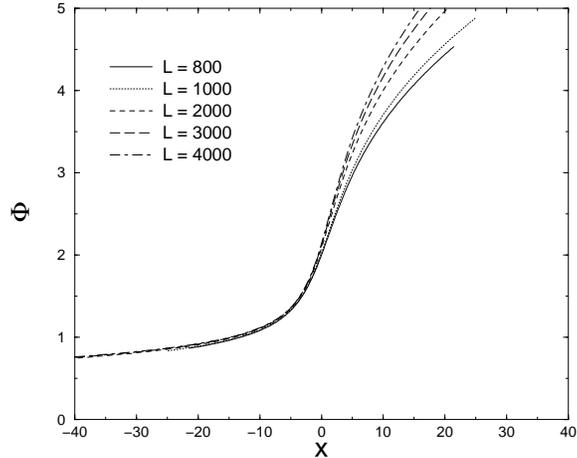}
\caption{Successive estimates for the scaling function $\Phi(x)$ are shown for increasing lengths of the walk.}
\label{scal_fig}
\end{figure}

\end{section}

\begin{section}{Conclusions}\label{conclusions}

In this work we have analyzed a model of self-avoiding partially directed chains on a square lattice, with a force pulling along one of the two lattice directions. The model is simple enough to yield the exact form of the temperature dependent critical force, $f_c(T)$ (see Eq. (\ref{critical_force})). However, the critical indices of the unfolding transition, which is second order, are not trivial. The transition is characterized by two exponents, the usual correlation length critical exponent $\nu$, and one which we called $\zeta$. In particular, the exponent $\nu$ at $f=f_c$ (see Eq. (\ref{scalfunc2})) is different both from $1/2$, the collapsed polymer value, and from $1$, the extended polymer value (see again Eq. (\ref{scalfunc2})). The $\zeta$ exponent characterize the singular behaviour of the chain elongation per monomer along the force direction as the critical force is approached from above (see Eq. (\ref{zeta_def})). Through a powerful enumeration technique taken from the literature \cite{prellberg}, coupled with a finite size scaling to extrapolate our results to infinitely long chains, we find that $\nu$ is very close to $2/3$. A scaling analysis also suggests that $\nu=\zeta$ at least within our numerical precision. Further investigations are required to extend our results to the undirected SAW case.

\end{section}

\end{document}